\documentclass[fleqn,usenatbib,useAMS]{mnras}

\usepackage{xcolor}
\usepackage{comment}
\usepackage{amsmath}
\definecolor{color}{RGB}{25,25,112}
\definecolor{negro}{RGB}{0,0,0}
\definecolor{colorurl}{RGB}{25,25,112}
 \usepackage{orcidlink}
\usepackage{amssymb}
\usepackage{subcaption}

\usepackage[T1]{fontenc}

\DeclareRobustCommand{\VAN}[3]{#2}
\let\VANthebibliography\thebibliography
\def\thebibliography{\DeclareRobustCommand{\VAN}[3]{##3}\VANthebibliography}

\usepackage{graphicx}	% Including figure files
\title[Degeneracy in GRB data analysis with MCMC]{Unraveling Parameter Degeneracy in GRB Data Analysis}

\author[Garcia-Cifuentes et al. 2023]{
Keneth~Garcia-Cifuentes\,\orcidlink{0009-0001-2607-6359},$^{1}$
Rosa~Leticia~Becerra\,\orcidlink{0000-0002-0216-3415},$^{1}$
Fabio~De~Colle\,\orcidlink{0000-0002-3137-4633},$^{1}$
and
Felipe~Vargas\,\orcidlink{0000-0001-5518-9689},$^{1}$
\\
$^1$ Instituto de Ciencias Nucleares, Universidad Nacional Aut\'onoma de M\'exico, Apartado Postal 70-264, 04510 M\'exico, CDMX, Mexico\\}
\begin{document}
\label{firstpage}
\pagerange{\pageref{firstpage}--\pageref{lastpage}}
\maketitle

\begin{abstract}
Gamma-ray burst (GRB) afterglow light curves and spectra provide information about the density of the environment, the energy of the explosion, the properties of the particle acceleration process, and the structure of the decelerating jet. Due to the large number of parameters involved, the model can present a certain degree of parameter degeneracy.
In this paper, we generated
synthetic photometric data points using a standard GRB afterglow model and fit them using the Markov Chain Monte Carlo (MCMC) method. This method has emerged as the preferred approach for analysing and interpreting data in astronomy. We show that, depending on the choice of priors, the parameter degeneracy can go unnoticed by the MCMC method. Furthermore, we apply the MCMC method to analyse the GRB~170817A afterglow. We find that there is a complete degeneracy between the energy of the explosion $E$, the density of the environment $n$, and the microphysical parameters describing the particle acceleration process (e.g. $\epsilon_e$ and $\epsilon_B$), which cannot be determined by the afterglow light curve alone.
Our results emphasise the importance of gaining a deep understanding of the degeneracy properties which can be present in GRB afterglows models, as well as the limitations of the MCMC method. 
In the case of GRB~170817, we get the following values for the physical parameters: $E=8\times 10^{50}-1 \times 10^{53}$ erg, $n=7\times 10^{-5}-9\times10^{-3}$, $\epsilon_e=10^{-3}-0.3$, $\epsilon_B=10^{-10}-0.3$.
\end{abstract}


%%%%%%%%%%%%%%%%%%%%%%%%%%%%%%%%%%%%%%%%%%%%%%%%%%

%%%%%%%%%%%%%%%%% BODY OF PAPER %%%%%%%%%%%%%%%%%%

\section{Introduction}

Gamma-ray bursts (GRBs) are generated during the propagation of highly relativistic jets \citep[see, e.g.,][for a review]{Kumar2015}. The gamma-ray emission results from the dissipation of kinetic or magnetic energy within the jet channel, while the subsequent multiwavelength afterglow is a consequence of the interaction between the relativistic jet and the surrounding medium \citep[][]{meszaros1997}. 

The afterglow phase can be generally\footnote{Very high-energy emission from GRB~190829A \citep{HESS2021190829A}, GRB~190114C \citep{Fermi190114C} and GRB~180720B \citep{180720B} are signatures of an inverse Compton component after the prompt emission.} explained as synchrotron radiation from a population of electrons accelerated by the relativistic shocks associated with the decelerating jet \citep[see, e.g.][for a recent review]{Miceli2022}.
In the most basic model of a spherical explosion propagating through a uniform medium, the Lorentz factor and velocity of the blast wave go as $\Gamma_{\rm sh} \propto t^{-3/2}$ during the relativistic phase, and $v_{\rm sh} \propto t^{-3/2}$ when the blast wave becomes sub-relativistic, being $t$ the time in the laboratory frame \citep{Blandford1976}.
Then, the afterglow emission is typically characterised by a series of power-law segments, with an observed flux density given as $F_{\nu}\propto t^{-\alpha}\nu^{-\beta}$, being $t$ and $\nu$ the time and frequency in the observer frame, and $\alpha$ and $\beta$ the temporal and spectral characteristic indices \citep[e.g.,][]{Sari1998, Granot2002, Becerra2019c,Becerra2021,Becerra2023b}.

By modelling the afterglow phase of GRBs, we can gain insight into the structure and dynamics of the relativistic jets produced during the burst, determine the structure of the surrounding medium, and potentially clarify the particle acceleration process responsible for the creation of the non-thermal population of electrons emitting synchrotron radiation. Detailed models of fitting have been presented for many GRB afterglows 
\citep[e.g.,][]{Tanvir2018,Bright2019,Cunningham2020,Fraija2020,OConnor2021,Gupta2022,Jelinek2022,Kumar2022,Salafia2022,Wang2022,Zhang2022,Angulo2023,Becerra2023b,Caballero2023,Laskar2023,Ren2023,deWet2023,Hussenot2023}, specially for  GRB~170817A, which rich multiwavelength dataset has driven numerous studies \citep[][]{Kim2017,Alexander2018,Granot2018,Lyman2018,Margutti2018,Resmi2018,Troja2018,Wu2018,Fraija2019,Gill2019,Hajela2019,Lamb2019,Troja2019,Ryan2020,Takahashi2020,Makhathini2021,McDowell2023,Ryan2023}.

Markov chain Monte Carlo (MCMC) methods \citep[see e.g.][]{Sharma2017,Hogg2018,Speagle2019}, in particular, are the most widely used strategies currently employed to model GRB observations.
These techniques have been used in the GRB community to determine not only multifrequency adjustments \citep[e.g.,][]{Dobie2018,Mooley2018b,Bright2019,Makhathini2021,Jelinek2022}, but also to model various components of the GRB, such as forward shock (most of the works cited above employ MCMC), reverse shock \citep[e.g.,][]{Laskar2016, Laskar2019} as well as prompt emission \citep{Li2022,Lazzati2023,Li2023}.
 based on the optimisation of the search for suitable parameters sampling probability distribution functions (PDFs).

% Parameters that can be restricted 
Parameters can be constrained by modelling the afterglow include the jet energy $E_j$, the jet opening angle $\theta_j$, and the density $n_\mathrm{0}$ of the circumstellar medium, in addition to microphysical parameters of the particle acceleration process, including the fraction of thermal energy in the population of accelerated electrons $\epsilon_e$ and in the magnetic field $\epsilon_B$, and the power law index of the population of non-thermal electrons $p$ \citep[see, e.g.,][]{Sari1998, Granot2002}. In addition, the emission depends on the jet structure and the observer angle, specially for observers located off-axis \citep[see, e.g.,][for a review]{Salafia2022b}.  

Given the large number of parameters involved in the model, one of the main challenges present when fitting GRB data using models is the possible presence of degeneracy in the parameter estimation. The issue of degeneracy has been recognised as a potential challenge in several articles \citep{Kim2017,Granot2018,Resmi2018,Troja2018,Wu2018,Fraija2019,Gill2019,Hajela2019,Laskar2019,Beniami2020,Cunningham2020,Angulo2023,Laskar2023,McDowell2023}, but its effects on the predictions of MCMC methods have not been discussed in detail.
Parameter degeneracy occurs when different sets of parameters reproduce the same set of system's behaviour. This can happen not only in the obvious case in which there are more parameters than observations, but also in the more subtle (and often difficult to identify) case in which several observations provide the same information, all together.

In this article, we examine both degenerate and non-degenerate models to assess the reliability of results obtained through the use of the MCMC method. Additionally, we apply it to the analysis of the GRB 170817A afterglow emission, as a special case where the degeneracy is highly evident. We will show that, if not considered carefully, parameter degeneracy can go undetected when applying MCMC techniques, and lead to potential misleading interpretations of the results. We stress, anyway, that the issue of parameter degeneracy is not limited to the MCMC algorithm or GRB 170817A itself, and it is common to most optimisation methods. Therefore, the same procedure and discussion presented here can be extended to any area where the number of parameters in the model is larger than the number of \emph{observables}, which subsequently leads to the emergence of degeneracies.

Our paper is organised as follows. In Section~\ref{sec:model}, we present a model of the afterglow phase of GRBs. In Section~\ref{sec:170817A}, we analyse in detail the results obtained by the application of MCMC to GRB~170817A.
Finally, in Section~\ref{sec:discussion}, we discuss the advantages and drawbacks of the MCMC method, the implications for the analysis of GRB~170817A, and summarise our conclusions.

\section{Fitting GRB afterglow data}
\label{sec:model}

%%%%%%%%%%%%%%%%%%%%%%%%%%%%
\begin{figure}
\includegraphics[width=\linewidth]{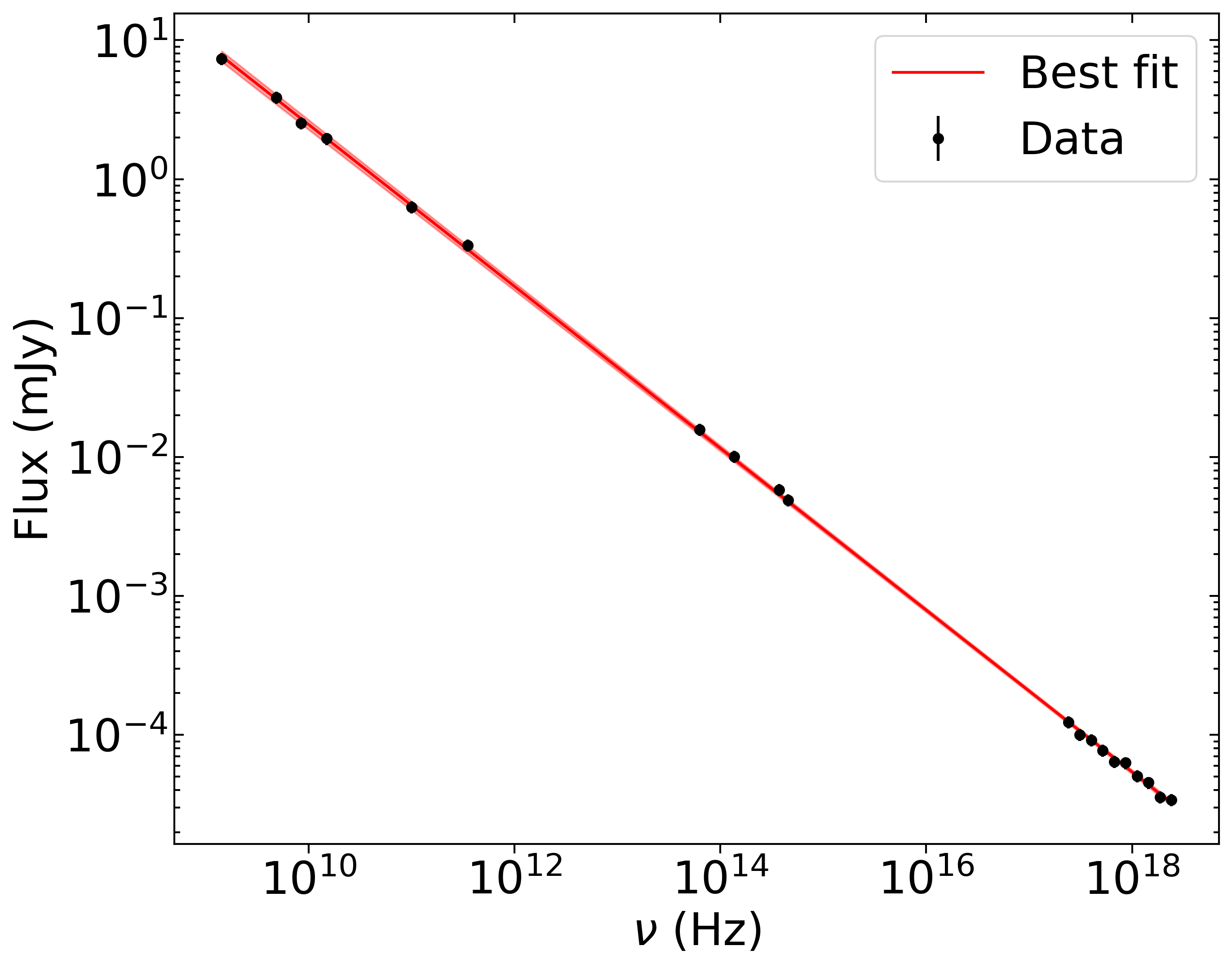}
  \caption{Best fit (red line) obtained using MCMC with the model function specified by equation \ref{eq:general_case}. Synthetic data points (black points) are generated using the same model, adding a Gaussian with Gaussian noise at a 10\% level.}
  \label{fig:GRBfit1}
\end{figure}
%%%%%%%%%%%%%%%%%%%%%%%%%%%%

\subsection{Analytical Model}

% Añadir la sección del fit que acabo de hacer

\label{sec:analytical}
In this section, we present the result of fitting degenerate models using MCMC. We consider as an example a simplified model of the GRB emission during the afterglow phase. In the next section, we show that the same problem is present when interpreting observations of the GRB170817A, among others.

GRB afterglows are typically modelled by considering synchrotron radiation from a population of electrons accelerated by shocks \citep[e.g.][]{Miceli2022}. The dynamics of a decelerating spherical blast wave is fully determined by considering its isotropic energy ($E_0$) and the density ($n$) of the environment. The synchrotron emission is typically described as a function of the microphysical parameters $\epsilon_e$, $\epsilon_B$ and $p$. These parameters give the fraction of thermal energy, magnetic energy and the power-law index of the electron energy distribution for the shock.

In general, the emission also depends on the characteristics of the emitting region, i.e. the jet opening angle ($\theta_{\rm j}$) in the case of a top-hat jet, or the full jet structure (specified by the energy $E(\theta)$ or shock Lorentz factor $\Gamma_{\rm sh}(\theta)$) in the case of a structured jet. Here, $\theta$ represents the polar angle, which is measured from the direction of the main axis of the jet. The function $E(\theta)$ is usually defined (in analytical models) by considering a specific angular dependence, that is, top-hat (where $E(\theta)=E_0$ when $\theta<\theta_j$ and zero otherwise, being $\theta_j$ the jet opening angle), Gaussian, or power-law jet structures \citep[see e.g.][]{Kumar2003,Beniami2019,Beniami2020}, among others, or calculated directly from the results of numerical simulations \citep[see e.g.][]{Lazzati2018,Xie2018,Salafia2020,Gottlieb2021,Urrutia2023}. Setting the jet structure substantially reduces the number of independent parameters. Furthermore, for jets observed off-axis (as the GRB~170817A), or when modelling observations made after the jet break time, it becomes necessary to also consider the position of the observer, specified by the observer angle ($\theta_{\rm obs}$).
 
The normalisation, slope, and general behaviour of the spectrum at a particular time depend on the characteristic frequencies. Inspired by observations of GRB~170817A, we consider a spectrum that lies entirely within the same spectral range at the specific time considered. 
In particular, we consider the case
$\nu_\mathrm{m}\ll \nu\ll \nu_\mathrm{c}$, where
the frequencies $\nu_\mathrm{m}$ and $\nu_\mathrm{c}$ correspond to the minimum and maximum energy of the electron population \citep[e.g.,][]{Sari1998, Granot2002}.
The time evolution of the flux at a given frequency depends on the details of the jet dynamics and the observer angle. For the dynamics described by the \citet{Blandford1976} self-similar evolution for an on-axis observer in an homogeneous environment, in particular, we have \citep{Granot2002}:
%In this spectral range\footnote{The same considerations apply to other spectral slopes, even though they yield different scaling for both the light curves and the spectrum \citep{Granot2002}.}, the emission is given as {\bf FABIO: es la emision o solo es un factor de rescalamiento? Check!}
%
\begin{eqnarray}
    F(\nu,t) = \alpha  n^{1/2} E_0^{(3+p)/4} \epsilon_e^{p-1} \epsilon_B^{\frac{p+1}{4}} \nu^{\frac{1-p}{2}} t_{\rm obs}^{3(1-p)/4}\;.
\label{eq:general_case}
\end{eqnarray}
Following \citet{Granot2002}, we indicate this spectral range as the PLS (power-law segment) ``G''. 
 
To see the impact of employing an MCMC fitting procedure on a degenerate problem, we further simplify equation \ref{eq:general_case}, by writing it as
\begin{eqnarray}
    F(\nu,~p,~\epsilon_\mathrm{e}) = k~\epsilon_\mathrm{e}^{p-1}~\nu^{\frac{1-p}{2}}, \quad k = g(E,n,\epsilon_\mathrm{B}, t_{\rm obs})
    \label{eq:order2}
\end{eqnarray}
Then, we set
\begin{eqnarray}
f(k,\epsilon_\mathrm{e}) = k \epsilon_e^{p-1} = 1.51\times10^6 {\; \rm (cgs\; units)}
\label{eq:f}
\end{eqnarray}
and we fit synthetic data generated using
equation \ref{eq:order2}. We set the constant $1.51\times10^6$ to obtain typical optical fluxes, corresponding to AB magnitudes between 16 and 22 \cite[see e.g.][]{Becerra2023}, and to explicitly show that $k$ and $\epsilon_\mathrm{e}$ are degenerate.

Figure~\ref{fig:GRBfit1} shows the synthetic spectrum generated by sampling data across radio frequencies ($\nu=1.43-35$ GHz in our example), optical ($\nu\sim 1$~eV), and X-ray frequencies ($\nu \sim 1$ KeV), and adding Gaussian noise at a 10\% level.
It is evident that, given the observed (synthetic) data points and the model described by the equation \ref{eq:order2}, the parameters $k$ and $\epsilon_e$ remain strongly degenerate. In fact, while the equation \ref{eq:order2} depends on three parameters (in addition to frequency), only two independent parameters can be determined from the synthetic spectrum, i.e. the slope of the spectrum (which implies that $p$ will be well constrained by the fitting procedure) and the flux normalisation $f$.

%%%%%%%%%%%%%%%%%%%%%%%%%%%%
\begin{figure}
\includegraphics[width=\linewidth]{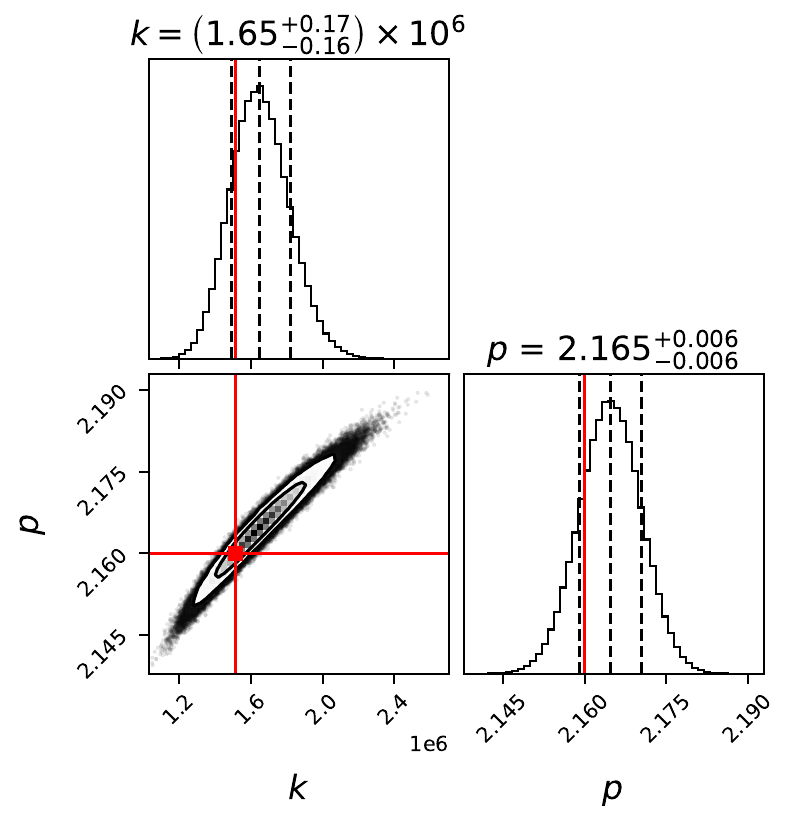}
 \includegraphics[width=0.95\linewidth]{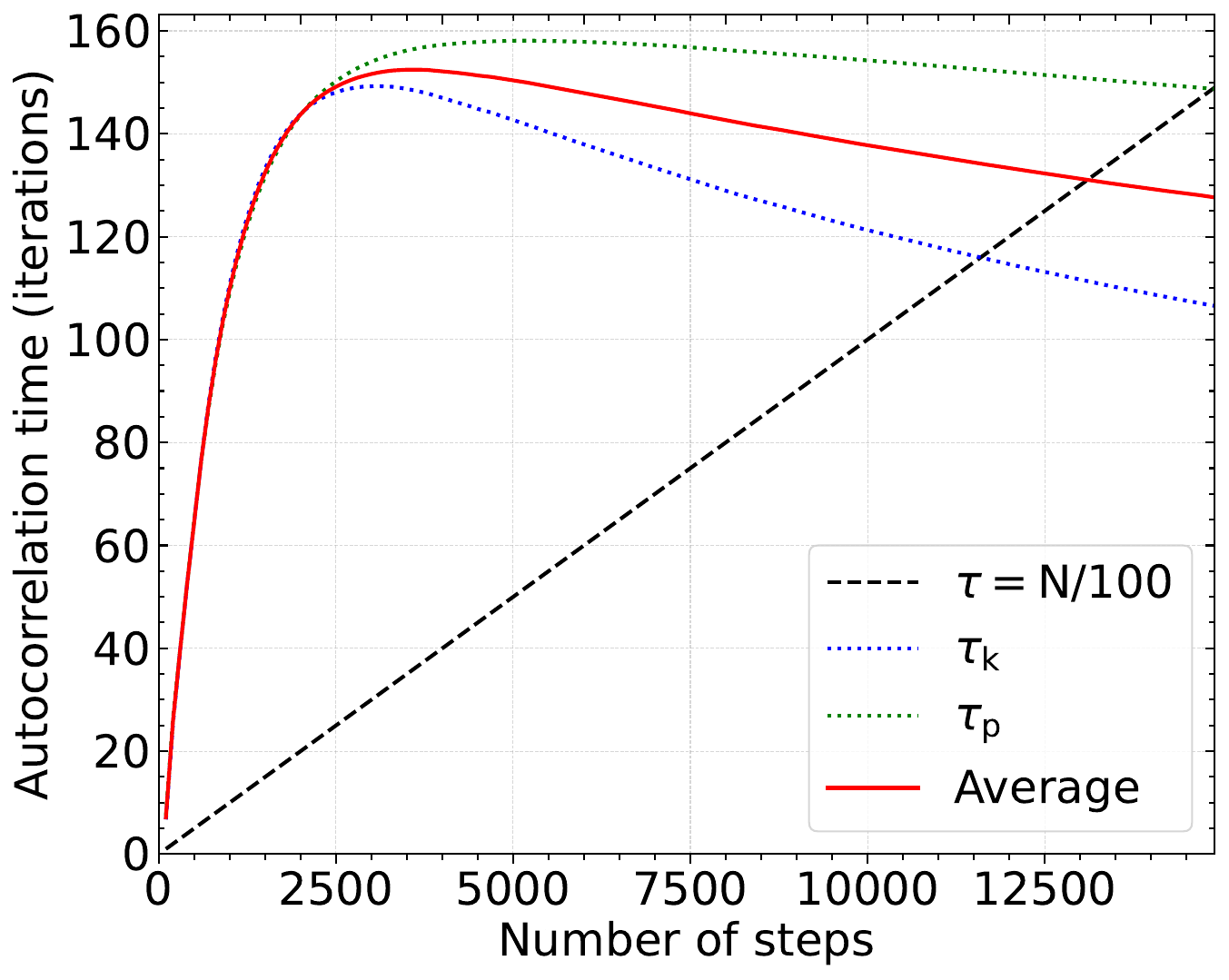}
  \caption{\emph{Top panel:} Corner plot obtained in MCMC on the non-degenerated model function (equation \ref{eq:order1}) using {\sc emcee} \citep{Foreman2013}. Labels $k,~p$ correspond to the fitted coefficients in the model. The values used to generate the synthetic data are shown with red lines.
   \emph{Bottom panel:} Autocorrelation plot obtained for the MCMC routine of the model in equation \ref{eq:order2}. The black dashed line represents the threshold $N = 100n$, with $n$ the number of iterations of MCMC. The red line represents the autocorrelation time $\tau$ estimated by the {\sc emcee}, and the dotted lines represent the autocorrelation time estimated on the parameters $k$ (blue line) and $p$ (green line) as subindices of $\tau$. }
  \label{fig:GRBfit1corner}
\end{figure}
%%%%%%%%%%%%%%%%%%%%%%%%%%%%

\subsection{Implementation of the MCMC method}
\label{sec:impl}

We perform the MCMC fitting using the Python package {\sc emcee} \citep{Foreman2013,Foreman2019} and the guidelines from Appendix~\ref{sec:mcmc}.

To check that the MCMC method provides the correct result for a non-degenerate model, we first fit our synthetic data points by using the non-degenerated model described by:
\begin{eqnarray}
    F(\nu,~p) = k~\nu^{\frac{1-p}{2}}\;, 
    %\quad k \propto f(E,n,\epsilon_\mathrm{e},\epsilon_\mathrm{B})
    \label{eq:order1}
\end{eqnarray}
where the parameters to be fitted correspond to a constant $k$ and $p$. %Then, we add a single-order degeneracy by including the fraction of thermal energy $\epsilon_\mathrm{e}$ in equation \ref{eq:order1}:
%\begin{eqnarray}
%    F(\nu,~p,~\epsilon_\mathrm{e}) = k~\epsilon_\mathrm{e}^{p-1}~\nu^{\frac{1-p}{2}}, \quad k \propto f(E,n,\epsilon_\mathrm{B})
%    \label{eq:order2}
%\end{eqnarray}
%Thus, there are three parameters to be fitted ($k$, $p$, and $\epsilon_\mathrm{e}$), and, in theory, there are an infinite number of combinations of the factors $k$ and $\epsilon_e$, part  of the product $k \epsilon_\mathrm{e}^{p-1}$, that can give a result equivalent to the constant $k$ defined in equation~\ref{eq:order1}.

We impose the conditions $2 \leq p \leq 3$ \citep{Achterberg2001, Shen2006,Levan2018} and $k > 0$ as priors for the log-prior function. We present the best-fit results (see Figure~\ref{fig:GRBfit1}), the corner plot obtained for the MCMC procedure (see the top panel of Figure~\ref{fig:GRBfit1corner}), and the autocorrelation plot (see the bottom panel of Figure~\ref{fig:GRBfit1corner}). The autocorrelation time provides an estimate of the maximum number of initial iterations that should be discarded before the system reaches equilibrium \citep{Sokal1997, Foreman2013}.

As expected, the results exhibit robust convergence, with the best fit values closely approximating the original parameters used to generate the synthetic data points, consistently falling within a confidence interval $1\sigma$.
The value of $k$ is determined with an error of $10\%$, consistent with the level of noise imposed on the data, while the value of $p$ has a much smaller uncertainty ($\sim 0.3\%$), due to the wide frequency range considered.

As can be seen in the bottom panel of Figure~\ref{fig:GRBfit1corner}, the autocorrelation values for both $p$ and $k$, followed over a span of 14200 iterations, fall below the $100N$ threshold imposed (see Appendix~\ref{sec:mcmc}). Therefore, we can state that, at the end of the iterative process, the Markov Chains related to the MCMC routine are independent of the initial position selected in the optimisation process, and the best-fit values obtained do not depend on the initial conditions.

\begin{figure}
\includegraphics[width=\linewidth]{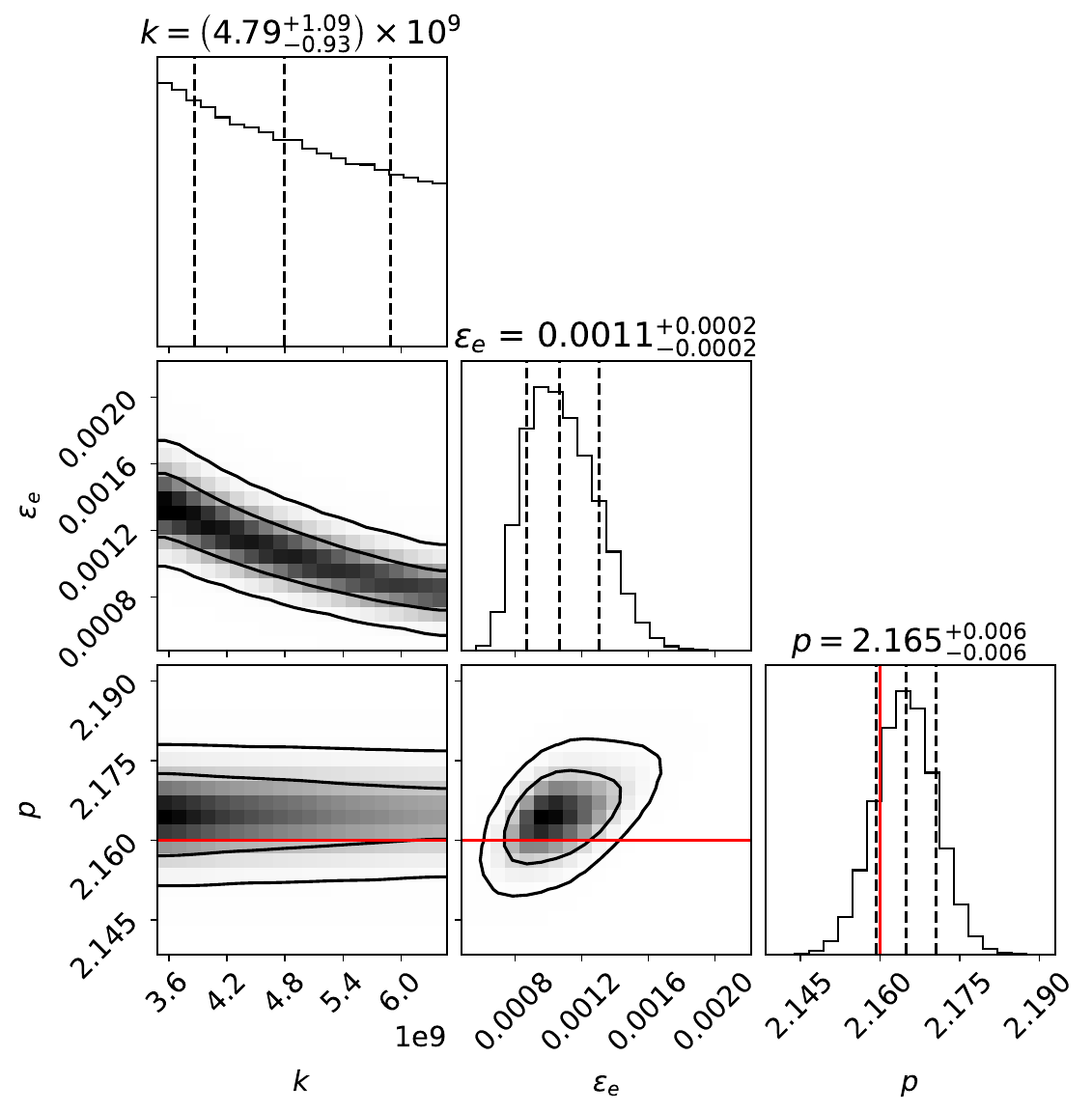} 
\includegraphics[width=.95\linewidth]{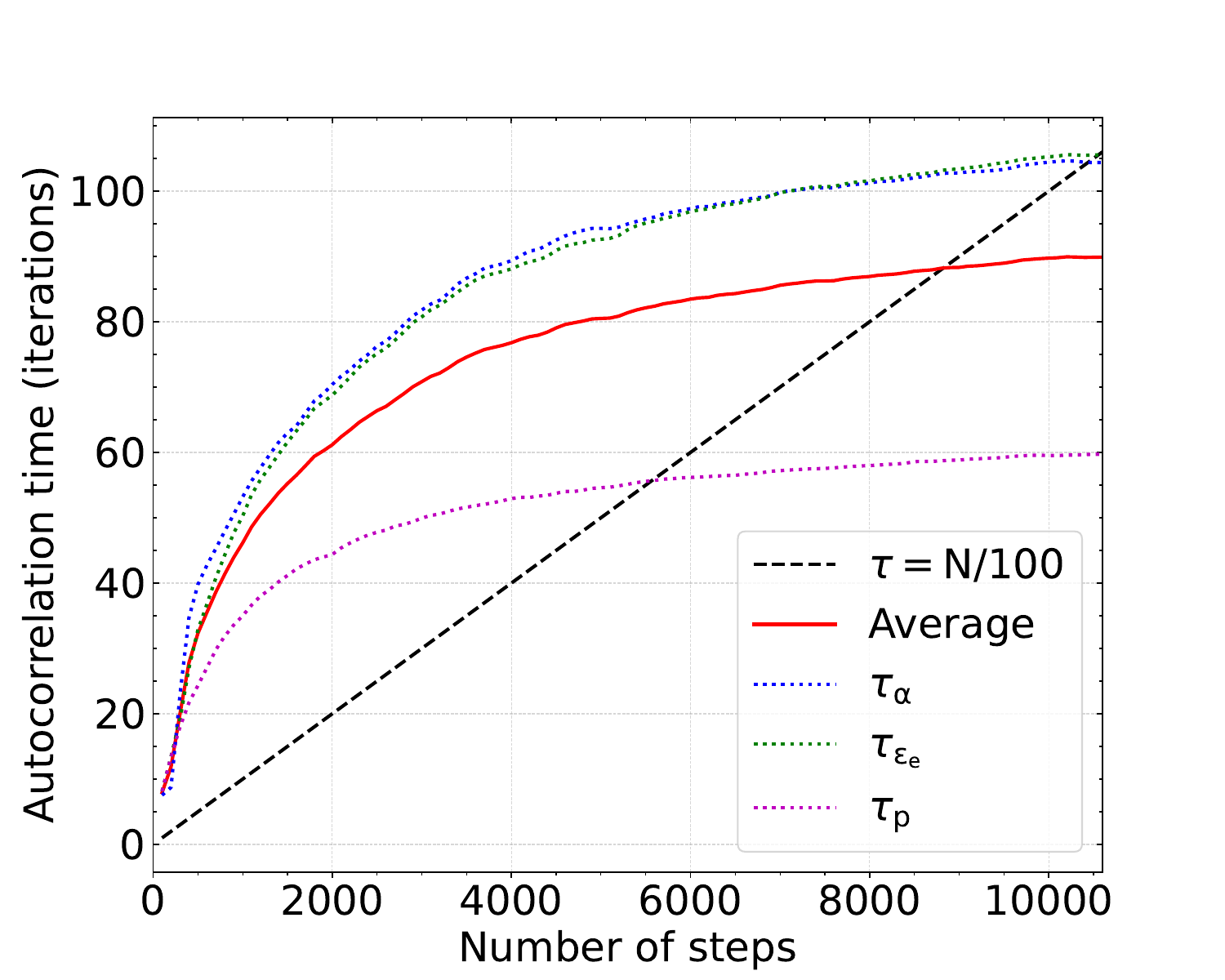}
  \caption{
\emph{Top panel:} Corner plot obtained in MCMC on the degenerated model function (equation~\ref{eq:order2}) using {\sc emcee} \citep{Foreman2013}. Labels $k, \epsilon_\mathrm{e},~p$ correspond to the fitted coefficients in the model of Eq.~\ref{eq:order2}. The original values provided for the synthetic data are shown with red lines.
\emph{Bottom panel:} Autocorrelation plot obtained for the MCMC routine of the model in equation~\ref{eq:order2}. The black dashed line represents the threshold $N = 100n$, with $n$ the number of iterations of MCMC. The red line represents the average autocorrelation time $\tau$ estimated by {\sc emcee}. The dotted lines represent the autocorrelation time for the parameters $\alpha$ (blue line), $\epsilon_\mathrm{e}$ (green line) and $p$ (magenta line).}
     \label{fig:GRBfit_new}
\end{figure}

Analogously, Figure~\ref{fig:GRBfit_new} presents the results of the MCMC fitting procedure obtained in the degenerate case (i.e., with the model specified by equation~\ref{eq:order2}). As in Figure~\ref{fig:GRBfit1corner}, we present here the corner plot (top panel) and the autocorrelation values estimated for each parameter and their average (bottom panel). For the priors, we impose the conditions $2 \leq p \leq 3$ \citep[e.g.,][]{Achterberg2001,Shen2006,Levan2018}, $10^{-5} \leq \epsilon_e \leq 0.3$ \citep{Beniamini2017}, and $3.5 \times 10^{9} \leq k \leq 6.5 \times 10^{9}$\footnote{This constraint is based on the combination of typical values in $n$, $E_0$ \citep{Berger2014}, and $\epsilon_\mathrm{B}$ \citep{Santana14} with the normalisation constant $\alpha$ from equation \ref{eq:general_case}.}. 
We use a logarithmic sampling, although similar results are obtained when using a linear sampling.

The MCMC procedure satisfies the autocorrelation criteria after $10600$ steps (see the bottom panel of Figure~\ref{fig:GRBfit_new}). The corner plot shows that the slope $p$ is well constrained ($p=2.165\pm 0.006$), and actually identical to the one shown in Figure~\ref{fig:GRBfit1corner}.
Although $k$ is poorly determined and the distribution of its value does not show a clear peak in the contour levels (see the top panel of Figure~\ref{fig:GRBfit_new}), $\epsilon_\mathrm{e}$ is well constrained ($\epsilon_\mathrm{e} = \left(1.1 \pm 0.2\right)\times 10^{-3}$).
This result is clearly incorrect and would lead, in a real physical problem, to a wrong estimation of the value of this parameter. Due to condition
$3.5 \times 10^{9} \leq k \leq 6.5 \times 10^{9}$ and the degeneracy of the model, all values $0.00074<\epsilon_e<0.00126$ are equally probable, as long as the product $k \epsilon_\mathrm{e}^{p-1}$ remains constant. Therefore, the distribution of the values of $\epsilon_e$ and $k$ should be flat in the range considered. An example of the correct flat distribution that should be obtained in a degenerate problem is shown, for example, in the middle panel of Figure~\ref{fig:toymodel}.

In summary, achieving a well-defined ``peaked'' distribution for a specific variable does not necessarily guarantee the precision of the estimation of this particular parameter, at least when dealing with degenerated systems.
When modelling simple physical systems, choosing a range of prior values too narrow can result in erroneous parameter estimations. In the next section, we show that the same problem is present when considering more complex degenerate problems, regardless of the width of the selected priors. 

\section{Fitting the GRB~170817A}
\label{sec:170817A}

\begin{figure*}
     \centering
\includegraphics[width=.95\textwidth]{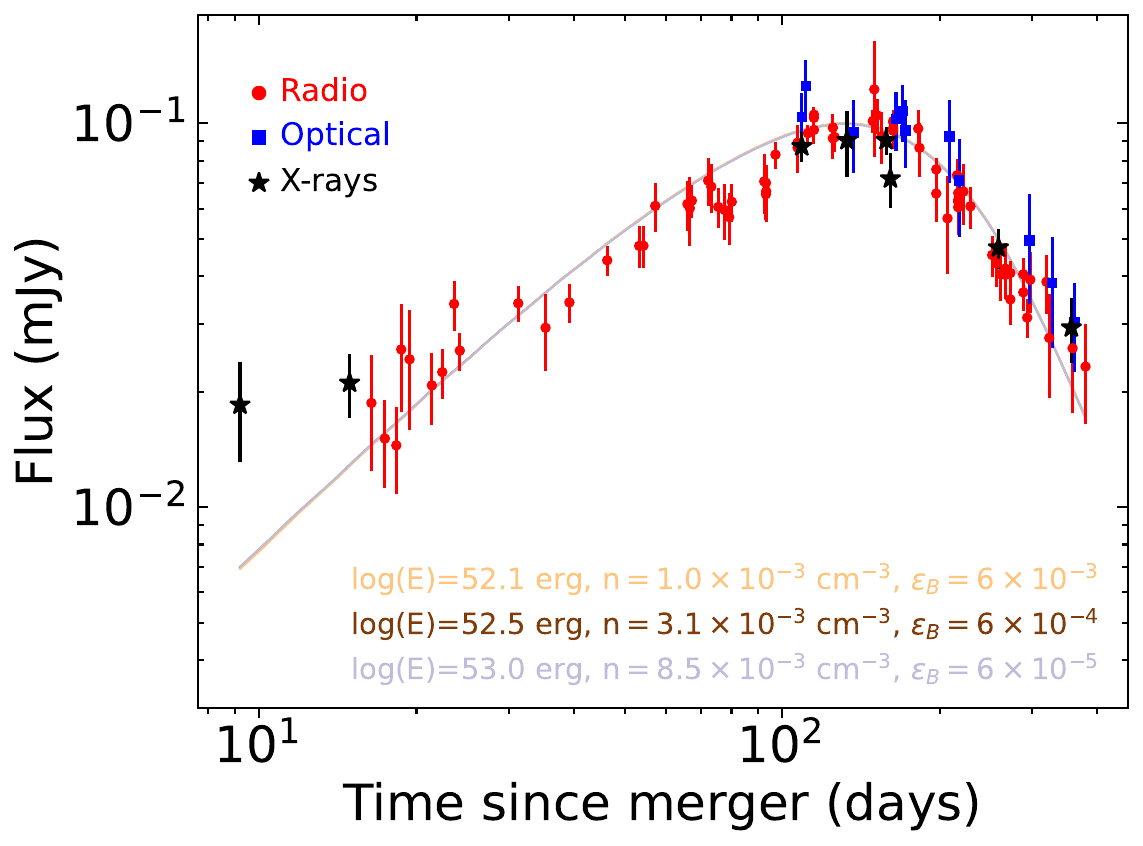}
  \caption{Fit of the GRB~170817A afterglow. The data points represent the observations, rescaled at $\nu_{\rm obs}=3\times 10^9$ Hz. The models are obtained for different combinations of the isotropic energy of the jet $E$, the density $n$ of the environment, and the microphysical parameter $\epsilon_E$, and illustrate the intrinsic degeneracy in the fitting process. Data taken from \citet{Makhathini2021}. 
  \label{fig:170817}
}
\end{figure*}

\begin{figure*}
  \includegraphics[width=.95\textwidth]{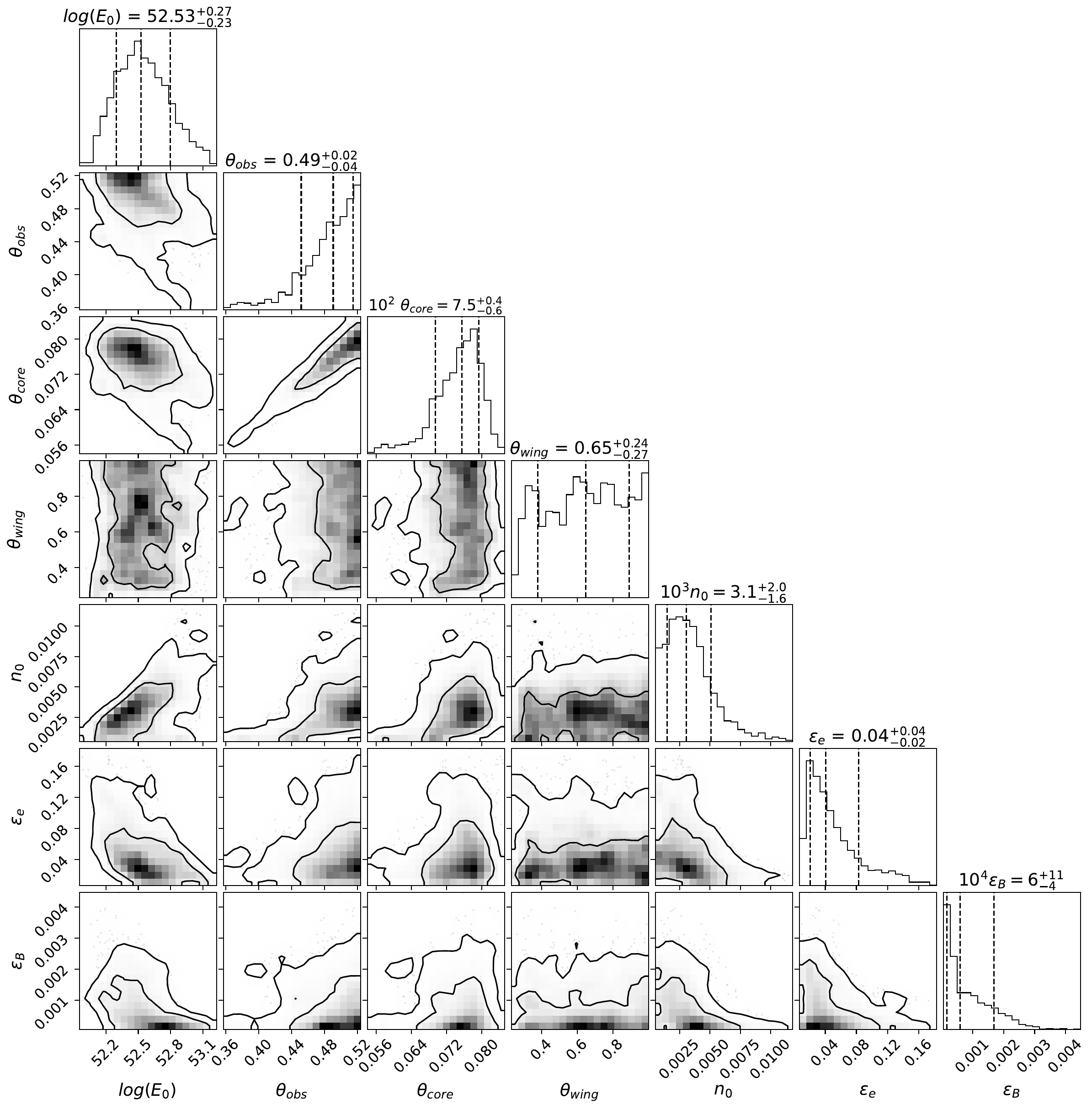}
  \caption{Corner plot of our fit of GRB~170817A afterglow. The values of the best fit obtained for each parameter are coloured red and listed in Table~\ref{tab:170817}. Data were taken from \citet{Makhathini2021}.}
  \label{fig:corner170817}
\end{figure*}

\begin{figure}
    \includegraphics[width=1\linewidth]{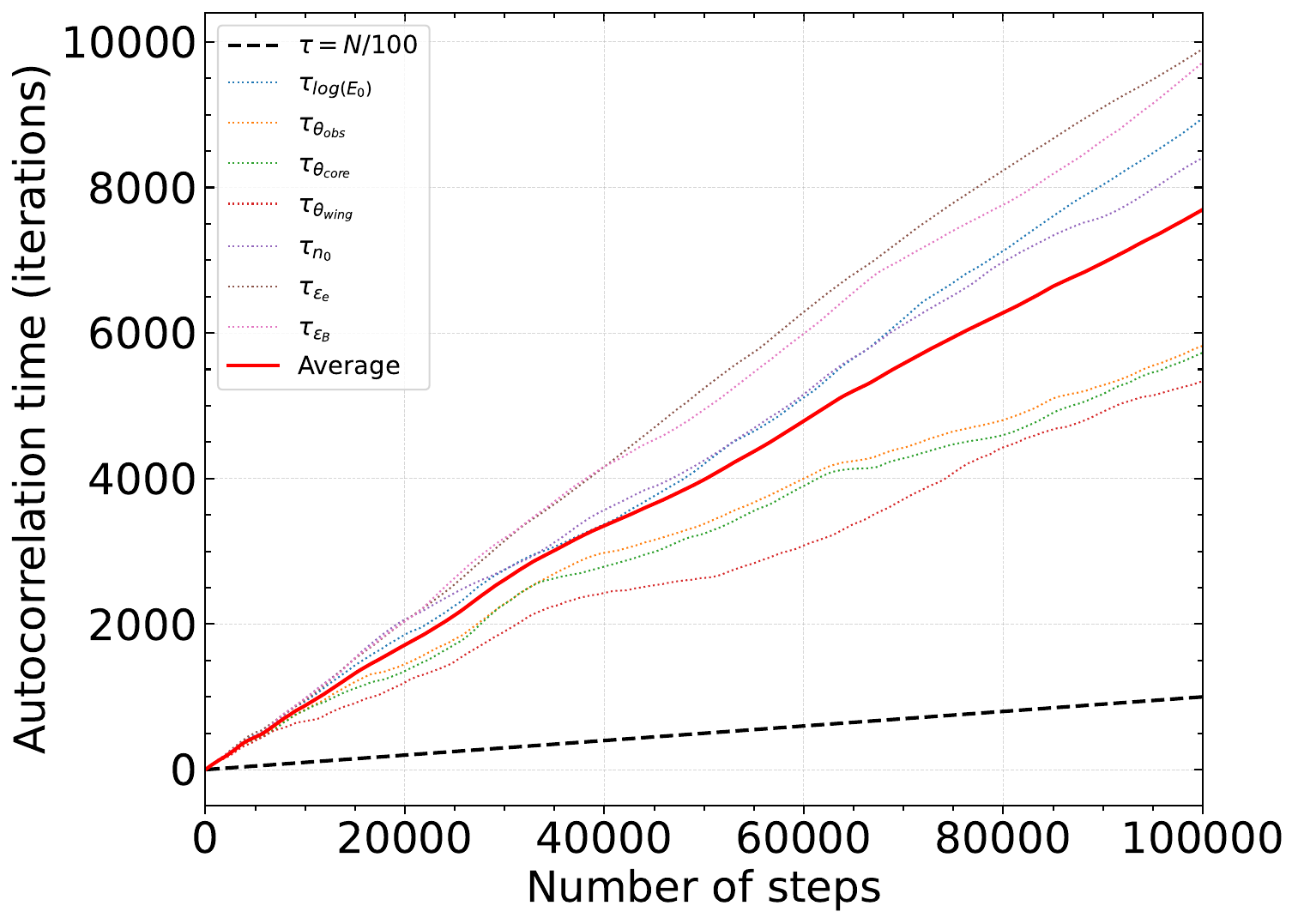}
    \caption{Autocorrelation plot of our fit of GRB~170817A afterglow showed in Figure~\ref{fig:corner170817}. This figure illustrates that there is no convergence for any of the parameters.}
    \label{fig:tau_afterglowpy}
\end{figure}

GRB~170817A, a low-luminosity short GRB, was detected by the {\itshape Fermi} and {\itshape INTEGRAL} telescopes beginning 1.7~s after the  gravitational wave (GW) signal observed by the LIGO and Virgo interferometers on 17 August 2017 \citep{Abbot2017,Abbot2017b}. Both events are the result of the merger of two neutron stars in NGC~4993, a lenticular galaxy located at a distance of 40~Mpc. The afterglow emission from the GRB~170817A can be successfully interpreted as the emission of a structured jet observed off-axis \citep[see, e.g.][]{DAvanzo2018,Margutti2018,Troja2018}.

\citet{Makhathini2021} presented the full multiwavelength afterglow light-curve data of GRB~170817A between 0.5 and 940~days after the merger, compiling all available photometry\footnote{Detections in X-ray started 9 days after the event. Between $T+0.5$ and $T+9$~days, the available photometry corresponds to upper limits that constrain the physical parameters very marginally.}.
\citet{Makhathini2021} fit the available data using a smoothly broken power-law model
\begin{equation}
    F(t,\nu) = 2^{1/s} F_p \left(\frac{\nu}{\rm 3\; GHz}\right)^{-\beta}\left[\left(\frac{t}{t_p}\right)^{-s\alpha_1}+\left(\frac{t}{t_p}\right)^{-s\alpha_2}  \right]^{-1/s}\;.
    \label{eq:mak}
\end{equation}
The six parameters of their model ($F_p$, $s$, $\alpha_1$, $\alpha_2$, $t_p$ and $\beta$) are determined precisely by using MCMC (see their figure 3). 
Consistent with this phenomenological model, the GRB~170817A spectrum falls, at all times, in the PLS G considered in the previous section \citep{Margutti2017,Troja2019}.
Thus, the parameter $\beta$ can be taken as constant as a function of time.  By fitting equation~\ref{eq:mak}, \citet{Makhathini2021} determined a value $\beta = 0.584 \pm 0.002$, corresponding to a slope $p = 2.168$ of the energy distribution of the accelerated electrons (being $\beta=(p-1)/2$). The data presented by \citet{Makhathini2021} are shown in Figure~\ref{fig:170817}, rescaled at the common frequency $\nu = 3$~GHz. 

Indeed, it is easy to see that just six parameters can be determined from the observations. These correspond to the slope of the spectrum $\beta$, the slope of the light curve as a function of time before and after the peak ($\alpha_1$ and $\alpha_2$ in equation \ref{eq:mak}), the flux normalisation $F_p$, the time $t_p$ corresponding to the peak in the flux, and the slope $s$ of the light curve close to its maximum. As this model fits well the available data, other models which include extra parameters (e.g., including a more complex description of the rise or fall of the light curve) will lead to poorly constrained or degenerated parameters. Furthermore, flux normalisation depends on $\epsilon_e$, $\epsilon_B$, $E_0$, and $n$. Thus, we expect that there will be a complete degeneracy between these parameters. This is illustrated in detail in the following. 

Figure~\ref{fig:170817} shows our fit of the photometry observations of GRB~170817A with the {\sc afterglowpy} library \citep{Ryan2020}. 
Figure \ref{fig:corner170817} shows the corner plot of our fit of the GRB 170817A afterglow, and Figure~\ref{fig:tau_afterglowpy} shows the evolution of the autocorrelation function for all parameters and their average. The values of the parameters inferred from the fit are shown in Table~\ref{tab:170817}. 
The model depends on the isotropic energy of the jet $E$, the density of the circumstellar medium $n$ (computed assuming a uniform medium), the microphysical parameters associated with the particle acceleration process $\epsilon_e$, $\epsilon_B$, $p$, the observer angle $\theta_{\rm obs}$, and parameters which determine the jet structure. We consider a Gaussian jet model, which depends on $\theta_{\rm core}$ and $\theta_{\rm wing}$, respectively, the angle at which most of the jet energy is contained and the maximum angle of the jet. As mentioned above, we fix $p=2.168$ following \citet{Makhathini2021}. As the value of $p$ is very well determined by spectral observations ranging from radio to X-ray frequencies, fixing it has the only effect of making the convergence of the fitting process faster and does not affect the results.
We also fix the fraction of accelerated electrons $\chi=1$ to limit the number of parameters.
The priors were selected for a wide ranges of values, i.e. $51 < \log\left(E_0\right) < 55$, $\theta_{\rm obs} < 30^{\circ}$, $\theta_{\rm core} < 20^{\circ}$, $\theta_{\rm wing} < 57.3^{\circ}$, $10^{-4} < n_0 < 1$ \citep{Berger2014}, and $10^{-6} < \epsilon_\mathrm{e, b} < \mathbf{1}$ \citep{Santana14}, with $E_0$ in ergs and $n_0$ in cm$^{-3}$.

\begin{comment}
    logE0_prior = (logE0 > 51) & (logE0 < 55)
    theta_obs_prior = (theta_obs > 0) & (theta_obs < 30*np.pi/180)  # 30 degrees
    theta_core_prior = (theta_core > 0) & (theta_core < 20*np.pi/180)  # 20 degrees
    theta_wing_prior = (theta_wing > 0) & (theta_wing < 1)
    n0_prior = (n0 > 1e-4) & (n0 < 1)
    epsilon_e_prior = (epsilon_e > 1e-6) & (epsilon_e < 1)
    epsilon_B_prior = (epsilon_B > 1e-6) & (epsilon_B < 1)
\end{comment}

The model fits the data satisfactorily (see Figure~\ref{fig:170817}). The corner plot illustrates whereas the different parameters are well constrained and how they are correlated with each other (by looking at the covariances between the different parameters).  
Figure~\ref{fig:corner170817} shows that some of the parameters are poorly constrained. For instance, the distribution of $\theta_{\rm wing}$ is nearly flat in the range $(0.4, 1)$. This is not surprising, as the energy at the edge of a Gaussian jet is very low. Thus, the emission from the jet at angles $\theta\lesssim \theta_{\rm wing}$ is weak and does not substantially modify the light curve. Although noisy, the covariance plots show that $\theta_{\rm wing}$ is not independent of other parameters. We can conclude that there is some level of degeneracy between $\theta_{\rm wing}$ and the other parameters, and the estimation of $\theta_{\rm wing}=0.65^{+0.24}_{-0.27}$ is not reliable.

We now focus our attention on other parameters, which seem well determined by looking at the corner plot. 
\citet{Nakar2021} have discussed in detail the degeneracy between $\theta_{\rm obs}$ and $\theta_{\rm core}$, showing that different combinations of these parameters, together with the shock Lorentz factor (which depends on the isotropic energy and ambient density here) produce the same results. 
Figure~\ref{fig:corner170817} shows that there is a certain degree of correlation between the values of $\theta_j$ and $\theta_{\rm core}$, although they are both determined with a small error and do not appear to be fully degenerate from the figure.

The projection of the posterior probability distributions of energy and density also presents a clear peak. Furthermore, the covariance between these two parameters clearly restricts the range of possible values to a small region of the parameter space (around $E_0=10^{52.53}$~erg and $n=3.1\times 10^{-3}$~cm$^{-3}$).
Thus, we conclude by looking at the fit and the corner plot that density and energy are also well constrained and not degenerated.

The results shown in the figure~\ref{fig:170817} and discussed in the last two paragraphs are both incorrect and can lead to a wrong interpretation of the physical conditions of this GRB. To verify the degeneracy between energy and density, in particular, we show in Figure~\ref{fig:170817} the afterglow light curve produced by three {\sc afterglowpy} models with different values of $E_0$, $n_0$ and $\epsilon_B$. Indeed, the light curves look nearly identical. This is consistent with the scaling expected in the particular spectral range in which the GRB 170817A observations fall at all times. The peak flux and peak time, in fact, scale as \citep{vanEerten2012,Granot2012}:
\begin{eqnarray}
    \frac{F_p}{F_{p,0}} &=& \left(\frac{E}{E_0}\right)\left(\frac{n}{n_0}\right)^{(p+1)/4} 
    \left(\frac{\epsilon_B}{\epsilon_{B,0}}\right)^{(p+1)/4} 
    \left(\frac{\epsilon_e}{\epsilon_{e,0}}\right)^{p-1}\;, \\
   \frac{t_p}{t_{p,0}} &=& \left( \frac{E}{E_0} \right)^{1/3} \left( \frac{n}{n_0} \right)^{-1/3} \;.
   \label{eq:rescaling}
\end{eqnarray}
That is, any combination of $E$, $n$, $\epsilon_e$ and $\epsilon_B$ that maintains unchanged $t_p$ and $F_p$, will produce the same light curve. This can be seen in Figure~\ref{fig:170817}, in which we have multiplied/divided the best value of energy and density by the same amount (a factor of three in this case) between the different models. 

As energy and density are rescaled by the same amount, equation \ref{eq:rescaling}
shows that, to maintain the same peak in flux, the scaling in energy must be related to the scaling in the microphysical parameters by the equation
\begin{equation}
  \frac{E}{E_0} = \left(\frac{\epsilon_e}{\epsilon_{e,0}}\right)^{-4(p-1)/(5+p)} \left(\frac{\epsilon_B}{\epsilon_{B,0}}\right)^{-(p+1)/(p+5)}\;.
  \label{eq:scaling}
\end{equation}
To get the parameters shown in Figure~\ref{fig:170817}, thus, we fix the value of $\epsilon_e$, we increase/drop $\epsilon_B$ by a factor of 10, and compute using this equation the resulting scaling in energy\footnote{The range of parameters producing a light curve identical to the one obtained in our best-fit model is much more extended than those indicated in Figure~\ref{fig:170817}, specially if the value of $\epsilon_e$ is lower.}. 
The scaling in density is taken equal to the one in energy, to maintain the position of the peak constant\footnote{Any scaling of $\epsilon_e, \epsilon_B, E$ satisfying equation \ref{eq:scaling} produce the same light curve as those shown in Figure~\ref{fig:170817} as long as the spectrum falls entirely (from radio to X-rays) in the PLS G, as the scaling depends on the spectral range considered \citep{Granot2012,vanEerten2012}.}. 

Figure~\ref{fig:tau_afterglowpy} shows the autocorrelation plot. As we can see, the autocorrelation plot does not show convergence either after 100 thousand steps. In this case, this plot implies that some of the parameters can be degenerated, and the results should be carefully considered. 

\begin{table}
\centering
\begin{tabular}{|l|c|c|}
\hline
\hline
Parameter & Value & Error \\
\hline
\hline
log($E_0$) & 52.5262& -0.2311 +0.2689\\
$\theta_\mathrm{obs}$ & 0.4905& -0.0388 +0.0243\\
$\theta_\mathrm{core}$ & 0.0754& -0.0060 +0.0039\\
$\theta_\mathrm{wing}$ & 0.6477& -0.2686 +0.2426\\
$n_0$ & 0.0031 &-0.0016 +0.0020\\
$\epsilon_\mathrm{e}$ & 0.0405& -0.0201 +0.0425\\
$\epsilon_\mathrm{B}$ & 0.0006& -0.0004 +0.0011\\
\hline
\end{tabular}
\caption{Parameters of the best fit model of GRB~170817A shown in Figure~\ref{fig:corner170817}.}
\label{tab:170817}
\end{table}

In conclusion, GRB~170817A, one of the most recently studied events, is an example of how a model of a light curve of a GRB using MCMC could lead to misleading results. Despite the large number of observations, the determination of the density, energy, and microphysical parameters for this event, as well as other parameters, remains uncertain.

\section{Discussion and Summary}
\label{sec:discussion}

%For any scientific model, it is necessary to make inferences about the world around us, that is, to help the said model to predict its behaviour. These assumptions can determine the probability distribution function of a particular event. 
In the previous section, we have shown that inferring parameters from a degenerate model leads to misleading results, which can go undetected by the MCMC algorithm.
Although a detailed analysis of the MCMC algorithm is outside the scope of this paper, as we do not pretend to present a full analysis of the advantages or drawbacks of this method, we have shown that:
a) the MCMC method correctly gives the correct solution in non-degenerate and simple degenerate problems.
b) When at least some of the priors are defined over a small range, degeneracy on other parameters can go unrecognised by the MCMC method (see section~\ref{sec:impl}). In this case, the MCMC may converge to a specific solution, instead of recognising the degeneracy present in the model.
c) The autocorrelation can be used as a criterion to determine whether a model is adequate or not \citep{Roy2020,BROOKS1998}, as long as the range of priors is not too narrow. However, as shown in section \ref{sec:impl},
%the previous section, 
the opposite is not true, and the convergence of the autocorrelation does not necessarily imply that the results converge to the correct solution. 

Therefore, in degenerated systems, a detailed analysis of the model is needed to understand which (if any) variable is degenerated, and to understand how many parameters can be well constrained by the data itself.

\subsection{How to break the degeneracy: the case of GRB 170817A}
\label{sec:break}

Given the number of parameters involved in models used to explain the phenomenology of GRBs, it is expected that in most cases there will be some degeneracy between the model parameters.
Without more observational data than photometry, it is very difficult to restrict the parameters of the model that describe the dynamics and emission properties of the system. 

%https://arxiv.org/pdf/1803.02768.pdf Resmi
%https://arxiv.org/pdf/1801.06516.pdf Troja

%\textbf{GRB 170817A is one of the most studied GRBs. The recompilation of the data set of this GRB is presented by \citet{Makhathini2021}. Before that work, using the observations carried out with the Giant Metrewave Radio Telescope \citet{Resmi2018} showed the distribution for the MCMC where the microphysical parameters were not restricted, identifying the degeneracy. \citet{Troja2018,Lyman2018,Wu2018} performed to model photometry from radio to X-rays in order to explain the structure observed in GRB 170817A. Nevertheless, the corner plots of the MCMC fit results presented in these works exhibited a degeneracy in several parameters.}

%\textbf{The recompilation of the data set of this GRB is presented by . Before that work, using the observations carried out with the Giant Metrewave Radio Telescope \citet{Resmi2018} showed the distribution for the MCMC where the microphysical parameters were not restricted, identifying the degeneracy. \citet{Troja2018,Lyman2018,Wu2018} performed to model photometry from radio to X-rays in order to explain the structure observed in GRB 170817A. Nevertheless, the corner plots of the MCMC fit results presented in these works exhibited a degeneracy in several parameters.}

Although fixing some of the parameters does not break the degeneracy, it helps to understand the relationship between the different parameters in the model. This is what has typically been done (implicitly) when comparing observations with numerical simulations. Relying on a set of underlying hypotheses on the energy, structure of the jet launched from the central engine, density of the dense environment (the star in the case of long GRBs or the debris of the neutron star merger for short GRBs), and so on, some of the parameters can be directly inferred from the result of the numerical simulation (e.g., the jet structure), and the observations can be then fitted, reducing the degree of degeneracy \citep[e.g.,][in the case of GRB 170817A]{Margutti2018,Wu2018,Xie2018}.

Although the problem of parameter degeneracy is often present when modelling GRB data, its relevance in limiting the parameters of the model is often underestimated when fitting data by using MCMC.
While a degenerate model should have as a solution a flat distribution of probability, we have shown that parameter degeneracy may not appear clearly in corner plots usually used to interpret the model fitting process.
This is the reason why, when the problem of degeneracy has been discussed in detail, at least in the context of GRB 170817A, this has been done mainly by starting from a theoretical understanding of the model, instead than starting from the result of the MCMC process.

\citet{Nakar2021} discussed the degeneracy between $\theta_{\rm core}$ and $\theta_{\rm obs}$, and, in general, the jet structure. They have reported different estimations present in the literature regarding these two parameters and have explained the differences as due to the different choice of priors. As we have shown, the convergence in the case of the afterglow fit of GRB 170817A can be extremely slow, which implies that, in fact, the final result can still depend on the initial choice of the parameters. This can be potentially verified by plotting the autocorrelation plot together with the corner plot.

GRBs seen on-axis typically provide limited information on the jet structure or on the observer angle, as most of the emission comes from the core of the jet, and it is weakly dependent on the emission coming from larger polar angles (but see, e.g., \citealt{Cunningham2020, Gill2023, OConnor2023} for cases in which the jet structure is determined from on-axis observations).
On the other hand, degeneracy in the model is particularly problematic in GRBs seen off-axis as the model will also depend on the jet structure. In the case of GRB~170817A, the full spectrum falls in the same spectral range (PLS G, corresponding to $\nu_\mathrm{m}\ll \nu\ll \nu_\mathrm{c}$), which implies that some of the parameters related with $\nu_m$ and $\nu_c$ ($\epsilon_e$, $\epsilon_b$, $E_0$, and $n_0$, see \citealt{Granot2002}) have a low value. Getting a spectrum that covers different power-law segments of the spectrum \citep[e.g.,][]{Harrison2001, Chandra2008, Perley2014, Tanvir2018, Salafia2022,Wang2022} can substantially, if not completely, reduce the degeneracy\footnote{We note that synchrotron emission remains intrinsically degenerate due to the uncertainty on the fraction of accelerated electrons, unless thermal emission can be used to infer the fraction of non-thermal over thermal electrons}. 

GRB 170817A is one of the most studied GRBs. \citet{Makhathini2021} presented the full data set, modelling the data with a phenomenological model. 
Their model includes six parameters, equal to the number of observables which can be inferred from the data set. Actually, most of the papers on GRB~170817A fitted more than six parameters. As shown in Section~\ref{sec:170817A}, there is a full degeneracy between $\epsilon_e$, $\epsilon_B$, $E$ and $n$.
Several authors \citep{Granot2018, Wu2018, Gill2019, McDowell2023} have discussed in detail the high level of degeneracy present in the model, considering the same scaling between the physical quantities discussed in this paper. 
\citet{Granot2018} concluded that the number of parameters outnumber the number of constraints in the data, and a significant level of degeneracy also remains when rich multiwavelength data are available, as in the case of GRB 170817A. 
\citet{Gill2019} also discussed in detail the degeneracy related to the scaling of the light curve. They determined a lower limit on the jet energy of $5.3\times10^{48}$~ergs, and on the ambient density of $5.3\times10^{-6}$~cm$^{-3}$, also considering observations of Very Long Baseline Interferometry (VLBI).
By considering the diffuse X-ray emission from the hot plasma in the host galaxy, \citep{Hajela2019} estimated an upper limit on the ambient density of $9.6\times 10^{-3}$~cm$^{-3}$. Using this estimation for the ambient density,  \citet{McDowell2023} estimated a jet energy of $7.5\times 10^{48}$~erg.
Using the value of the density estimated by \citep{Hajela2019} as an upper limit, instead, we get the following ranges for the physical parameters: $E=8\times 10^{50}-1 \times 10^{53}$ erg, $n=7\times 10^{-5}-9\times10^{-3}$, $\epsilon_e=10^{-3}-0.3$, $\epsilon_B=10^{-10}-0.3$.

The degeneracy can also be reduced when observations of polarisation are available, or the jet is resolved spatially, i.e. by VLBI observations. The upper limit of 12\% on the linear polarisation of GRB 170817A available at $244$~days in radio frequencies was used to constrain the magnetic field anisotropy factor $B_\parallel/B_\perp = 0.5-0.9$ both from analytical methods and numerical simulations \citep{Gill2020,Medina2023}, consistent with previous estimates \citep{Granot2003,Stringer2020}.
In principle, polarisation can also place constraints on the other parameters of the afterglow model. 
The same is true if VLBI observations are available (see, e.g., \citealt{Taylor2004, Mooley2018, Ghirlanda19, Salafia2022}). 

Another approach to reduce the degeneracy is to use a smaller subset of parameters that can be constrained without a global fit of the GRB light curves \citep[see, e.g.][]{Santana14,Beniamini2017}. For instance, as demonstrated by \citet{Beniamini2017}, it is possible to constrain $\epsilon_e$ by using the peaks of radio light curves.

Finally, we note that, in the era of multi-messenger astrophysics, simultaneous measurements of GRBs with gravitational waves and/or neutrinos allow search intervals to be eventually reduced \citep[e.g. see][]{Finstad2018,Lyman2018, Dichiara2021}.
%Estimating the constraints on the density, energy, and other parameters of the GRB 170817A due to polarisation, afterglow and VLBI proper motion is left to a future study. 

\subsection{Final thoughts}

In this paper, we have presented several examples to understand the effect of fitting the afterglow of GRBs using the MCMC method. The ease of implementation, as well as the availability of different free libraries, has made this method very popular in the analysis of astrophysical data. Despite the widespread use of this technique, the degeneracy eventually present in the model has not been thoroughly considered in the GRB literature.

In this work, we have emphasised the importance of considering the possible parameter degeneracy and implementing appropriate priors when interpreting the results obtained by employing the MCMC algorithm. 
Although the computational time required for correct modelling observations can be considerable \citep{South2022}, we highlighted the importance of conducting a convergence analysis, which is necessary to demonstrate that the derived interpretations are reliable \citep{Roy2020,BROOKS1998}. In cases of parameter degeneracy, the lack of convergence can be clearly observed in the autocorrelation plot (see Figure~\ref{fig:tau_afterglowpy}). This occurs because the MCMC can get trapped in a region of high probability density that may not necessarily represent the correct physical model.
It is also important to note that MCMC can converge if the initial range of priors has been chosen too narrow. In such cases, MCMC may find a solution that fits the data well, but in reality is constrained by the initial priors and may not accurately represent the correct physical model. 

Regarding GRB~170817A, we have shown that it is not possible to determine the energy of the explosion, the density of the environment and the microphysical parameters $\epsilon_e$ and $\epsilon_B$ related to the particle acceleration process from the afterglow light curve alone. Furthermore, for the specific case of GRB 170817A, the complete radio to X-ray afterglow spectrum lies on the same spectral segment making the degeneracy problem much more severe. Nevertheless, the degeneracy can be identified in most GRBs. a few examples include GRB 230812B \citep{Hussenot2023}, GRB 221009A \citep{Laskar2023}, GRB 210205A \citep{Gupta2022}, GRB 210104A \citep{Zhang2022}, or GRB 160625B \citep{Cunningham2020}.

% GRB most used in papers, then it is most usual to see degeneration problems in this GRB.

Dealing with parameter degeneracy is very difficult not only for the MCMC method but more generally for most optimisation methods. This highlights the importance of fully understanding the possible presence of parameter degeneracy in a model during the fitting process. There are several practical steps that can be taken to avoid misinterpreting the results of MCMC, including using multiple Markov chains using different priors\footnote{By doing this, we can diagnose if the chains converge to the same PDF.}, implement other diagnosis tools such as the Gelman and Rubin statistics \citep{Gelman92}, or by choosing priors that better reflect previous knowledge or assumptions about the parameters, rather than using flat priors. Moreover, it is recommended to use alternative fitting algorithms such as local and global minimizers, nested sampling or machine learning methods \citep{Kochoska2020}.

Finally, the degeneracy problem extends beyond the analysis of GRBs, and the discussion presented in this work is applicable to any other field in astrophysics that involves model fitting with a limited number of observables. In such cases, it is recommended to exercise particular caution with employing fitting techniques like MCMC, specially in situations where parameter degeneracy may remain unnoticed. Failing to detect degeneracy might result in misleading results, analysis and interpretations.

\section*{Acknowledgements}

The authors would like to thank the anonymous reviewer for his/her valuable comments.
The authors acknowledge César Fernández Ramírez for useful discussions. 
We acknowledge the support from the DGAPA/PAPIIT grants IG100422 and IN105921.
KGC and FV acknowledge the support of the CONAHCyT fellowship. RLB acknowledges support from the CONAHCyT postdoctoral fellowship. 
FDC acknowledges the computing time granted by DGTIC UNAM on the supercomputer Miztli (project LANCAD-UNAM-DGTIC-281).
%%%%%%%%%%%%%%%%%%%%%%%%%%%%%%%%%%%%%%%%%%%%%%%%%%

\section*{Data availability}
The data underlying this article will be shared on reasonable request to the corresponding author.


\bibliographystyle{mnras}
\bibliography{references} % if your bibtex file is called example.bib

\appendix

\section{MCMC: Markov Chain Monte Carlo}
\label{sec:mcmc}

Markov Chain Monte Carlo methods are algorithms used to draw random samples systematically from high-dimensional probability distributions. These approaches have been widely used in the astrophysics literature \citep[see][for a review]{Sharma2017}.

{\sc emcee} \citep{Foreman2013} is an actively developed open-source Python module that implements the Affine Invariant Markov Chain MCMC Ensemble sampler from \citet{Goodman10}, designed to run on multicore CPUs.

When using the {\sc emcee}, it is necessary to set the distributions for the log-prior, log-likelihood, and log-probability functions. The log-prior function reflects previous knowledge, assumptions or information about the parameters, and it imposes constraints on the parameters of a model \citep{MacKay2002}. The log-likelihood function calculates the acceptance probability for each step within the parameter space for every chain.
(i.e., it measures the accuracy of fitting a statistical model), and the log-probability function maps the parameter space and is used to calculate the PDF in bayesian inference \citep{MacKay2002, Foreman2013}.

In this work, to avoid making assumptions on the a priori distribution, we use flat priors on all parameters in every MCMC routine. Moreover, we use a like-chi square formula as log-likelihood function:
\begin{eqnarray}
    l(\theta) = \frac{1}{2} ~ \sum_i^N \frac{\left[F_i(t_i, \nu_i) - F(\theta, t_i, \nu_i)\right]^2}{F_{e, i}^2(t_i, \nu_i)}\;,
\end{eqnarray}
where $t_i$ and $\nu_i$ refer to the time and frequency data $i$, respectively, $F_{e,i}$ is the uncertainty associated with each value of the observed flux $F_{i}$, $F(\theta, t_{i},\nu_i)$ corresponds to the model function to be fitted to the data, and $\theta$ are the set of parameters associated (see section \ref{sec:analytical}). 

In MCMC approaches, one of the most important parameters is the number of \textit{workers} in the process. This parameter represents the number of parallel chains running simultaneously in the MCMC routine, which are used to explore the parameter space on the sampler \citep{Foreman2013}. In general, it is recommended to use hundreds of workers, although it is possible to go beyond (i.e., thousands of workers). In this paper we use 1000 workers to obtain a sufficient number of independent samples per autocorrelation time (whose value provides an estimate of the maximum number of initial iterations that should be discarded before the system reaches equilibrium \citep{Sokal1997, Foreman2013}.

A possible misuse of {\sc emcee} is to set a process with a fixed (and random) number of samples. As \citet{Foreman2013} recommended, we only need to run the sampler a few autocorrelation times. Beyond that limit, we obtain an independent set of samples. With that in mind, we monitor the progress of the chain using an autocorrelation time criterion $\tau$ that is updated every $100$~steps \citep{Foreman2013}. Therefore, we stop the sampler if the Markov chain has a length $L$ with $L \geq 100 \tau$ and if $\tau$ has not changed by more than $1\%$ with respect to the previous estimation of $\tau$ (100 iterations before). In this way, the convergence of the Markov chain is ensured, removing the need to use a fixed number of steps. However, we set a maximum number of iterations of $10^5$ to handle cases where there is no clear convergence (i.e., degenerate models with a very low convergence rate).

By employing an optimisation step before sampling can substantially mitigate the computational cost associated with posterior probability density functions, leading to more enhanced and well-suited fittings. Therefore, we use a maximum likelihood estimation before executing the MCMC sampling process. 

Moreover, it is important to note that MCMC is a sampler, not an optimiser \citep{Hogg2018}. Therefore, if an optimisation of the posterior probability density functions (PDFs) or likelihood is required, another process will be needed before implementing MCMC.

\section{Degenerated models}
\label{sec:toy}

\begin{figure}
\centering
      \includegraphics[width=0.85\linewidth]{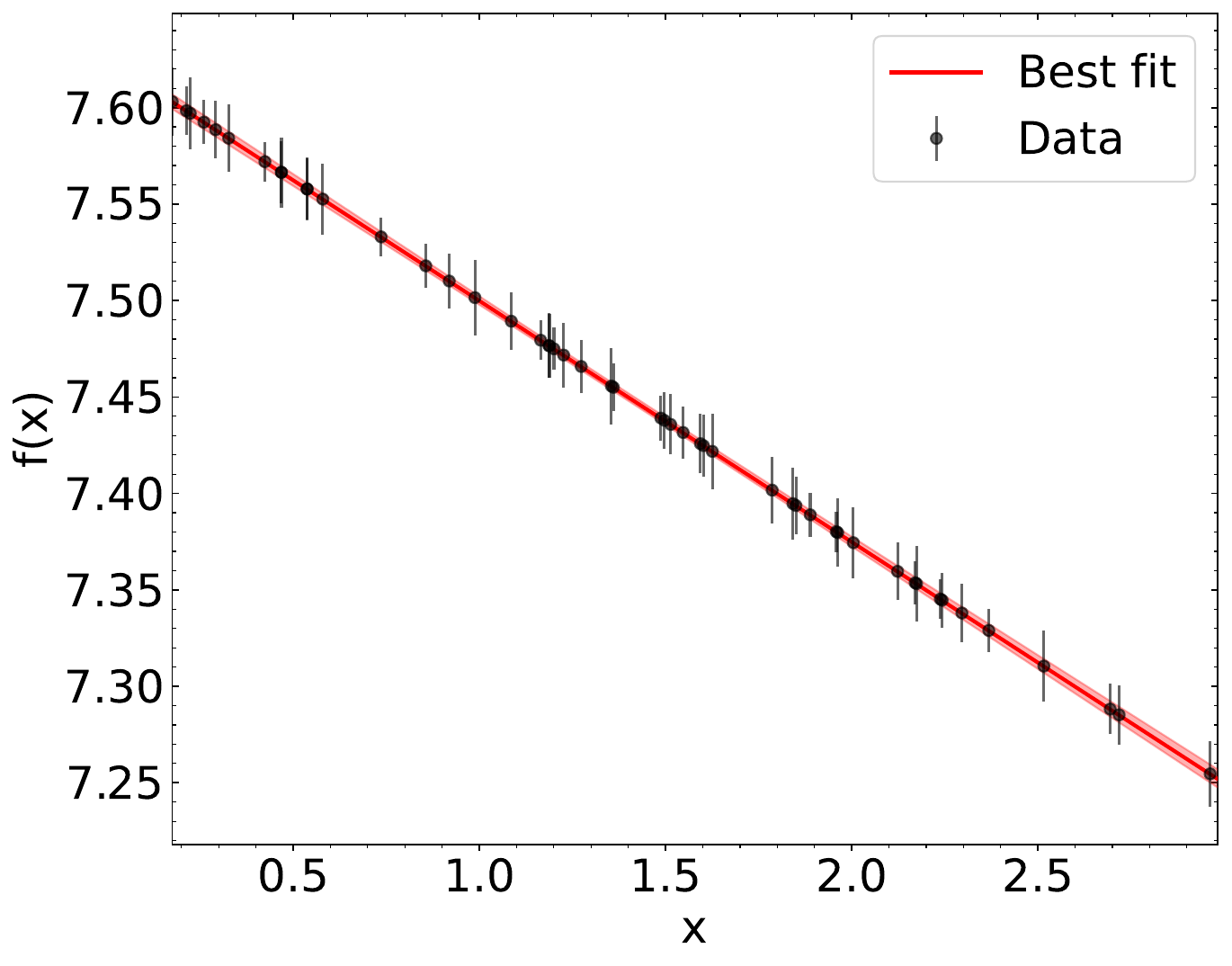}
    \includegraphics[width=0.85\linewidth]{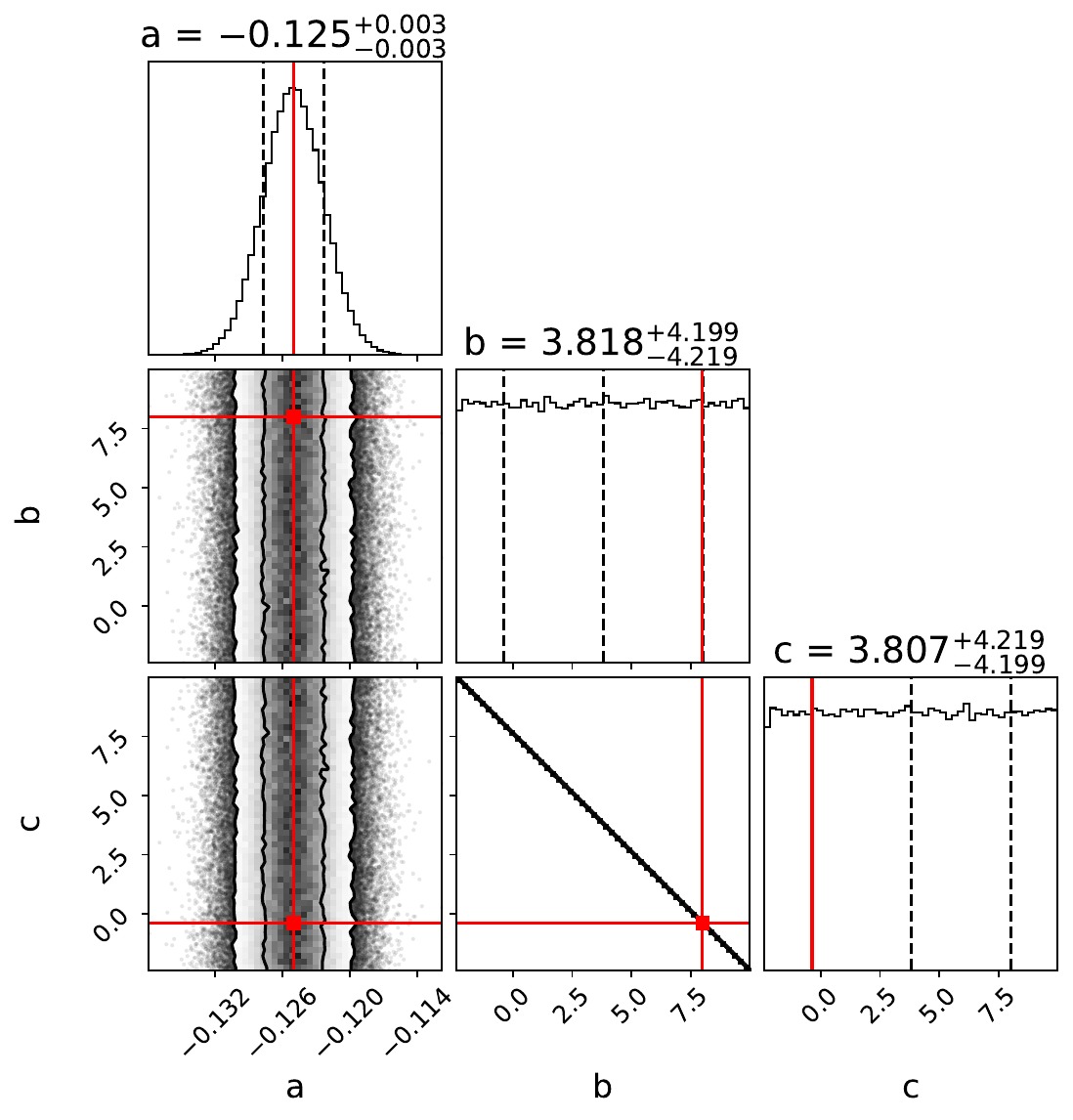}
      \includegraphics[width=0.81\linewidth]{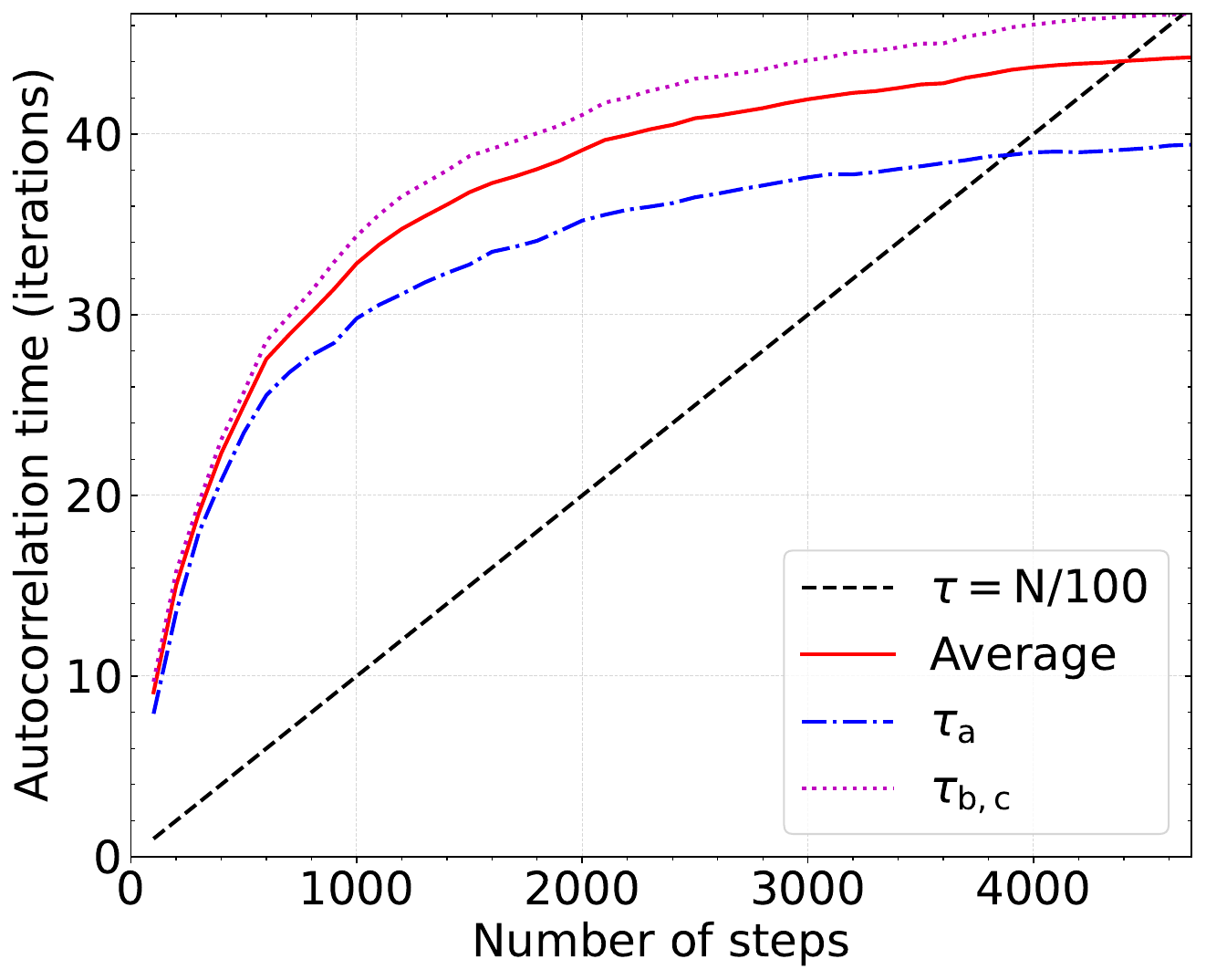}
  \caption{
     \emph{Top panel:} Synthetic data points (black lines) and best fit (red line) obtained using MCMC with the model function of equation \ref{eq:toymodel} of the data with Gaussian noise (black points). \emph{Central panel:} Corner plot obtained in MCMC on the degenerated model function (equation \ref{eq:toymodel}) using {\sc emcee} \citep{Foreman2013}. The labels $a,~b,~c$ correspond to the fitted coefficients in the model. The true values used to generate the synthetic data are shown with red lines. \emph{Bottom panel:} Autocorrelation plot obtained for the MCMC routine of the model in equation \ref{eq:order2}. The black dashed line represents the threshold $N = 100n$, with $n$ the number of MCMC iterations. The red line represents the average autocorrelation time $\tau$ estimated by {\sc emcee}. The magenta dotted line represent $\tau$ for parameters $b$ and $c$, and the blue dash-dotted line represent $\tau$ for parameter $a$.}
     \label{fig:toymodel}
\end{figure}

In order to show the result of a degenerate fitting using MCMC, we consider a simple model given by the line:
\begin{eqnarray}
    f(x) = -\frac{1}{8} x + \frac{61}{8}\;,
\end{eqnarray}
where the slope and constant values were selected arbitrarily. We add Gaussian noise with an amplitude of $0.01$ (see the top panel of Figure~\ref{fig:toymodel}). We assume that the fitting function has the following form:
\begin{eqnarray}
    g(x) = a x + (b + c)
    \label{eq:toymodel}
\end{eqnarray}
where $a,~b~$, and $c$ are the fitting parameters for the MCMC procedure. There are an infinite number of combinations for $b$ and $c$ satisfying the condition $b+c = 61/8$. Following the same procedure described in Appendix \ref{sec:mcmc}, we obtain the best fit results, the autocorrelation diagram, and the corner plot for the MCMC routine (Figure~\ref{fig:toymodel}). 

As can be seen in the middle panel of Figure~\ref{fig:toymodel}, the slope (a non-degenerate parameter) is well constrained, leading to the best fit value of $a = -0.1250 \pm -0.0027$. However, the parameters $b$ and $c$ remain (correctly) undetermined, and the convergence criteria have been satisfied at step $4700$ (see the bottom panel of Figure~\ref{fig:toymodel}). Furthermore, as can be seen in the bottom left panel of Figure~\ref{fig:toymodel}, the autocorrelation time related to the slope $a$ (blue dotted line) is lower than the autocorrelation time related to the parameters $b$ and $c$ (magenta dotted line) and therefore few iterations are needed to converge.

The histograms related to $b$ and $c$ are almost flat in the range of values considered by the sampler (see the middle panel of Figure~\ref{fig:toymodel}). Therefore, it can be assumed that every value between these ranges is equally likely, consistently with the condition $b+c = 61/8$. 
%%%%%%%%%%%%%%%%%%%%%%%%%%%%

% Don't change these lines
\bsp	% typesetting comment
\label{lastpage}
\end{document}